\documentclass{article}

\def\cP{\mathcal{P}}

\def\cC{\mathcal{C}}

\usepackage{amsmath,amssymb,amsthm,graphicx,verbatim,tabularx}
\usepackage{algorithmic,algorithm,newfloat,color,longtable}
\usepackage[hyphens]{url}

\DeclareFloatingEnvironment[name=Algorithm]{alg}

\begin{document}

\title{Comparing Voting Districts with
    Uncertain Data Envelopment Analysis}

\author{Casey Garner$^{a\dagger}$ and Allen Holder$^{\, b^*\dagger}$ \\ \\
\parbox{.9\textwidth}{\footnotesize
$^a\,$Department of Mathematics, University of Minnesota,
	Minneapolis, MN, USA, (garne214@umn.edu)} \\[8pt]
\parbox{.9\textwidth}{\footnotesize
$^b\,$Department of Mathematics, Rose-Hulman Institute of Technology,
        Terre Haute, IN, USA (holder@rose-hulman.edu)} \\[4pt]
\parbox{0.9\textwidth}{\footnotesize
$^*$Corresponding author} \\[0pt]
\parbox{.9\textwidth}{\footnotesize
$^\dagger\,$ authors listed alphabetically and contributed equally,
	research conducted at Rose-Hulman Institute of Technology} \\[8pt]
\parbox{.9\textwidth}{\footnotesize {\color{blue}
$\bullet$ This preprint has not undergone peer review (when applicable) or any post-submission improvements or corrections. The Version of Record of this article is published in {\it Optimization Letters}, and is available online at \url{https://doi.org/10.1007/s11590-023-02046-0
}}} \\[8pt]
}

\maketitle

\begin{abstract}
Gerrymandering voting districts is one of the most salient concerns 
of contemporary American society, and the creation of new voting
maps, along with their subsequent legal challenges, speaks for 
much of our modern political discourse. The legal, societal,
and political debate over serviceable voting districts demands a
concept of fairness, which is a loosely characterized, but amorphous,
concept that has evaded precise definition. We advance a new paradigm
to compare voting maps that avoids the pitfalls associated with
an a priori metric being used to uniformly assess maps. Our evaluative 
method instead shows how to use uncertain data envelopment analysis 
to assess maps on a variety of metrics, a tactic that permits each 
district to be assessed separately and optimally. We test our 
methodology on a collection of proposed and publicly available maps 
to illustrate our assessment strategy. \\[8pt]

\hspace*{54pt}Gerrymandering \\[4pt]

\end{abstract}

\section{Introduction and Strategy} 
	\label{sec-gerrymandering}

The question of how to draw congressional boundaries in the US is among the 
most politically charged of debates, and as such, it has drawn significant
attention in both our contemporary literature, 
see e.g.~\cite{astor2019,daley2017,mcgann2016,re_2019}, 
and in our academic literature, see~\cite{benade2020,bozkaya2003,chatterjee2019,%
king2012,king2015,king2018,ludden2019,pegden2017,ricca2013} as well as 
the related districting problems in~\cite{caro2004,damico2002,gentry2015}.
Recent concerns about partisan tactics have lead to legal 
cases~\cite{russell2020}, and gerrymandering has been linked to
economic impacts~\cite{akey2017}. Moreover, there are numerous 
amalgamations of federal and state laws that blur any sense of an ideal 
process, a fact that regularly promotes wrangling among political 
factions.

Our goal here is not to enter the highly contested fray of drawing 
congressional boundaries per se, and we are not advocating for any specific
type of map or for any particular map quality. We instead have the singular 
goal to demonstrate that it is reasonable to compare different congressional maps
relative to a collection of agreed upon characteristics.  We example
a couple of justifiable characteristics and use them to compute a
map's proximity to equitable efficiency~\cite{holder2022}, which 
is an evaluation that permits each map, and indeed each district, to
be assessed on its own optimal metric. The map closest to equitable efficiency 
is then declared to be the best of the comparative cohort.

We adopt the sense of efficiency from data envelopment analysis (DEA), which is 
a stalwart analytic technique that separates decision making units (DMUs) into 
those that are efficient and those that are not. There are numerous DEA 
adaptations, see~\cite{cooper2007} as a review, but we build 
from the recent work in~\cite{ehrgott2018,holder2022}, the latter of which
defines the proximity to equitable efficiency. We neither review DEA nor 
its recent robust/uncertain extensions for the sake of brevity, although 
readers can find recent reviews and advances 
in~\cite{peykani2020,salahi2021,TOLOO2022102583}. The proximity 
to equitable efficiency is the smallest amount of uncertainty required of the 
defining characteristics so that all DMUs have a claim to efficiency with 
an appropriate selection of data from within the collection of uncertain 
possibilities. So a collection of DMUs with a small proximity value indicates 
a shared efficiency status, whereas a large proximity value suggests disparities 
among the DMUs. We note that small proximity values do not imply similar 
characteristics but instead mean that each DMU has a way to weight its defining
characteristics to reach an efficiency score of one.

\subsection{Characteristic Data}

Congressional districts segment a state's population by assigning 
individuals to districts, and we assume that $C^s$ represents a district
and that $\cC = \{ C^s : s = 1,2,\ldots, S\}$ is a proposed map for a 
particular state.  We select two characteristics for each district, one 
input and one output, the intents of which are:
\begin{quote}
\begin{description}
\item[input-characteristic ---] minimize wasted votes, and
\item[output-characteristic ---] maximize proportional representation.
\end{description}
\end{quote}
We tout the use of one input and one output in our example
because the proximity value is restricted within a range whose
bounds have a ratio no greater than $\sqrt{2}$, which is the 
smallest possible ratio~\cite{holder2020}.
The use of only two characteristics further supports graphical 
interpretation.  However, our assessment strategy extends to an
arbitrary number of inputs and outputs, and political and social
scientists could straightforwardly include more and different 
characteristics.

Eight different maps are compared for each state from FiveThiryEight's
{\it The Atlas of Redistricting}.\footnote{\url{https://projects.fivethirtyeight.com/redistricting-maps/}}
Each district in the atlas associates with 1) a chance of being 
represented by a democrat, 2) a chance of being represented by a republican, 
and 3) the size of the voting age population, as well as several other 
statistics.  We further collect the political leaning of each state from 
the Pew Research Center,\footnote{\url{https://www.pewforum.org/religious-landscape-study/compare/party-affiliation/by/state/}} data that divides each state's population
into percentages that are, or lean, republican; are, or lean, democrat;
and that have no tendency. We simplify this information by evenly splitting 
the percentage having no tendency between the other two categories.
We use the following notation for these statistics:
\begin{center}
\begin{tabular}{rcl}
$\rho^D_{skt}$ & -- &
	chance that district $s$ in state $k$ \\
	& & under map $t$ goes democratic, \\[4pt]
$\rho^R_{skt}$ & -- &
	chance that district $s$ in state $k$ \\
	& & under map $t$ goes republican, \\[4pt]
$v_{skt}$ & -- & voting age population of district $s$ \\
	& & in state $k$ under map $t$, \\[4pt]
$\ell^D_k$ & -- &
	likely percentage to vote democratic \\
	& & in state $k$, and \\[4pt]
$\ell^R_k$ & -- &
	likely percentage to vote republican \\
	& & in state $k$. \\[6pt]
\end{tabular}
\end{center}
We assume two parties, and as such, $\rho^D_{skt} = 1 - \rho^R_{skt}$
and $\ell^D_k = 1 - \ell^R_k$.

Our input-characteristic is a normalized proxy for wasted votes and 
is modeled as,
\begin{eqnarray*}
X_{skt} 
& = & \frac{ \left( \rho^L_{skt} \, v_{skt} + 
	\left( \rho^W_{skt} \, v_{skt} 
		- \rho^L_{skt} \, v_{skt} - 1 \right) \right) -
	\frac{1}{2} \, v_{skt} } 
	{ \left( \rho^L_{skt} \, v_{skt} +
        \left( \rho^W_{skt} \, v_{skt}
                - \rho^L_{skt} \, v_{skt} - 1 \right) \right) + 
        \frac{1}{2} \, v_{skt} } \\[6pt]
& = & \frac{ \left( \rho^W_{skt} - \frac{1}{2} \right) \, v_{skt} - 1 } 
	{\left( \rho^W_{skt} + \frac{1}{2} \right) \, v_{skt} - 1},
\end{eqnarray*}
where 
\[
\rho^L_{skt} = \min \left\{ \rho^D_{skt}, \, \rho^R_{skt} \right\} 
	\; \mbox{ and } \;
\rho^W_{skt} = \max \left\{ \rho^D_{skt}, \, \rho^R_{skt} \right\}.
\]
We assume without loss of application that the size of the voting age 
population satisfies $v_{skt} \ge 1 / (\rho^W_{skt} - (1/2))$, which 
presupposes the example's reality that $\rho^W_{skt} > 1/2$, i.e. no district 
under any map has perfect parity with regard to its chance of being 
represented by either party.  This assumption ensures the nonnegativity
of $X_{skt}$.  As an illustration, suppose a district with 
$10000$ eligible voters has an $80\%$ chance of going republican and a 
$20\%$ chance of going democratic.  We interpret this information as there 
being a likely outcome of $8000$ republican votes and $2000$ democratic 
votes, in which case the losing party wasted all $2000$ votes and the 
winning party wasted $8000 - 2000 - 1= 5999$ votes.  The total number of 
wasted votes is thus $7999$.  The minimum number of wasted votes in a two 
party system is half of the total number of votes, and our numerator is 
$7999 - 5000 = 2999$ as a deviation from this ideal.  
The denominator normalizes this calculation and guarantees that 
\[
0 \le X_{skt} < \frac{1}{3} 
	\left( \frac{ v_{skt} - 2 } {v_{skt} - \frac{2}{3}} \right)
	\approx \frac{1}{3},
\]
where the approximation is asymptotically perfect as the population
grows. Voting age populations are in the hundreds of thousands, and
our practical upper bound on $X_{skt}$ is 1/3.  The input-characteristic 
for the example is $2999/12999 \approx 0.23$.  Notice that $X_{skt}$ is
near its smallest assumed value of zero only if 
\[
\rho^W_{skt} = \frac{1}{2} + \frac{1}{v_{skt}} \approx \frac{1}{2},
\]
which coincides with there being a minimum number of wasted votes
under our assumption on the size of the voting age population.

The output-characteristic represents each district's importance with
regard to statewide proportional representation. Let $Z_{skt}$ be a
binary random variable so that an observation of $1$ means that district $s$
is won by the anticipated party and an observation of $0$ means that 
the district is lost by the anticipated party.  The output-characteristic 
we assume is a conditional expectation of the form,
\[
Y_{skt} = E \left( \left. \frac{ \left| \gamma_k - \psi(Z_{kt}) \right| }
		{ \gamma_k + \psi(Z_{kt}) } \; \right| 
			\; Z_{skt} = 0 \right).
\]
The value of $\gamma_k$ is the targeted proportional representation for state
$k$, represented by
\[
\gamma_k = \frac{ \max \left\{ \ell_k^D, \, \ell_k^R \right\} }
           { \min \left\{ \ell_k^D, \, \ell_k^R \right\} }.
\]
The random vector $Z_{kt}$ indexes over $s$ and represents a possible outcome 
of an election in state $k$ under map $t$.  For instance, if state $k$ 
under map $t$ has $5$ districts, then an election outcome could be
\[
Z_{kt} = (1,0,1,1,0),
\]
which would mean that districts $1$, $3$, and $4$ were won by the 
anticipated party but that districts $2$ and $5$ were upset by the 
non-anticipated party. 

The function $\psi$ calculates the ratio of majority to minority party 
districts won. Suppose for instance that $Z_{kt} = (1,0,1,1,0)$ as above and 
that the Republicans hold a $60\%$ majority in the state.  Then the statewide 
Republican to Democrat ratio is $\gamma_k = 0.6/0.4 = 1.5$.  Suppose further 
that the first four districts are majority Republican and are thus anticipated 
to be won by Republicans.  The election represented by $Z_{kt}$ shows that 
Republicans won three of their four anticipated districts and that 
Democrats lost their one anticipated district.  The ratio of majority to 
minority district wins in this case is $4/1 = 4$, which for this
particular outcome of $Z_{kt}$ gives,
\[
\frac{ \left| \gamma_k - \psi(Z_{kt}) \right| }{ \gamma_k + \psi(Z_{kt}) }
	= \frac{|1.5 - 4|}{1.5 + 4} = 0.45.
\]
As a second illustration, suppose we are calculating the importance of 
the third district, which is majority Republican.  The most anticipated
outcome with district three being lost by the Republicans is 
$Z_{kt} = (1,1,0,1,1)$, and the ratio of majority to minority district wins 
in this case is $3/2 = 1.5$. The result for this outcome of $Z_{kt}$ is thus,
\[
\frac{ \left| \gamma_k - \psi(Z_{kt}) \right| }{ \gamma_k + \psi(Z_{kt}) }
	= \frac{|1.5 - 1.5|}{1.5 + 1.5} = 0.
\]
This calculation suggests that district three's importance toward achieving 
statewide district proportionality is diminished, especially if the
sample election $(1,1,0,1,1)$ is highly likely given the district's loss.  
Republican districts one, two, and four have similar interpretations in this 
case, although their $Y_{skt}$ values are unlikely to be identical due to 
their varying chances of being won or lost. We note that the denominator 
defining $Y_{skt}$ normalizes deviation from an assumed ideal
like the denominator of $X_{skt}$, which helps maintain scale.
We further note that the evaluation of $\psi(Z_{kt})$ could have been
justifiably weighted by the districts' populations, its just that we have 
instead decided to count districts as representative party units.

We simulate elections to estimate $Y_{skt}$.  One concern with calculating 
$Y_{skt}$ during a simulation is that the minority party might lose all
districts in any particular random election, which would make the denominator
of $\psi(Z_{kt})$ zero.  In this case we set $\psi(Z_{kt})$ to be the number 
of majority wins, which is a reasonable upper bound on all other cases --
one that guarantees a finite expectation.  Each $Y_{skt}$ is the average
of 1000 sample means, with each being based on 250 random elections.
A simulated election assigns each district other than $s$ an independent 
sample from a standard uniform variable, and $Z_{skt}$ is $1$ if the 
outcome is no more than $\max \left\{ \rho^D_{skt}, \, \rho^R_{skt} \right\}$.
Otherwise $Z_{skt}$ is $0$ and the district is won by the district's
minority party.

\subsection{Voting Maps}

Eight different congressional maps are considered for each state.
These maps are defined in {\it The Atlas of Redistricting}%
\footnote{\renewcommand\footnoterule{\rule{\linewidth}}%
\url{https://fivethirtyeight.com/features/we-drew-2568-congressional-districts-by-hand-heres-how/}} and are:
\begin{itemize}
\item {\bf Current (Crt)} -- the current congressional map (as of 2020),
\item {\bf Republican (Rep)} -- designed to favor Republicans,
\item {\bf Democrat (Dem)} -- designed to favor Democrats,
\item {\bf Ratio (Rto)} -- designed to match statewide partisan representation,
\item {\bf Competitive (Cpt)} -- designed to increase competitive elections,
\item {\bf MajMin (MMn)} -- attempts to maximize the number of majority-minority
			districts,
\item {\bf Compact (Cmt)} -- designed for most compact districts, and
\item {\bf County (Cty)} -- designed for compact districts that conform to 
	county borders.
\end{itemize}
The compact map is generated algorithmically\footnote{See Brian
Olson's BDistricting at \url{https://bdistricting.com/}}, but all
others are hand drawn by the author's of {\it The Atlas
Project}, with designs being guided by general principles like 
requiring contiguous districts.  We note that the term minority means
racial or ethnic minority, and not political minority, when used
to describe a map.

Several of the maps are defined by a deliberate objective, and some of
these objectives align with our inputs and outputs. 
Specifically, our suggestion to increase proportional representation
agrees with the intent of the ratio map, and our suggestion to decrease
wasted votes agrees with the intent of the competitive map. This observation 
might seem to bias our outcomes toward these maps, but this expectation
is fallacious against our results. The reason is that the proximity to
equitable efficiency is not solely defined by one of these objectives but is
instead defined by each district's best employ of the characteristics so
that it can simultaneously seek to maximize proportional representation 
and minimize the number of wasted votes -- doing so in a way that best
competes with the other districts of the map.  

\subsection{Experimental Commentary}

The input data of our model is without doubt uncertain due to constant
changes in demography and political sway. Selecting and promoting maps with 
a minimal proximity to equitable efficiency submits a preference toward maps
that can uniformly achieve efficiency scores of one across their districts
with only small adjustments in the political divisions imposed by
the map. So a map most proximal to equitable efficiency is closest to a 
sense of `perfect' fairness, by which we mean that each district has a way 
to measure itself with regard to the characteristics in a way that matches
the others. If the proximity value is zero, then the map achieves this 
sense of equity without adjustment.

A collection of matrices defines the structure of uncertainty, and we
assume identity matrices in our example for simplicity. We could adjust 
these matrices to explore different outcomes, but designing characteristics 
and detailing their relationships is generally the responsibility of experts 
should they use our proximity measure to evaluate maps. Moreover, 
our uncertainty assumption is mathematically on $X_{skt}$ and $Y_{skt}$,
but the lack of knowledge really stems from the imperfect nature
of the supporting data, i.e. from $\rho^D_{skt}$, $\rho^R_{skt}$, 
$v_{skt}$, $\ell^D_k$, and $\ell^R_k$.

Each map associates with a heuristically estimated 
$\sigma^* = (\sigma_X^*, \, \sigma_Y^*)$ that expresses (near) 
minimum levels of uncertainty required of the transformed data $X_{skt}$ and  $Y_{skt}$,
and the proximity to equitable efficiency is $\cP(\cC) = \| \sigma^* \|$.
The search for $\sigma^*$ occurs within a region satisfying 
$\| \hat{\sigma} \| / \sqrt{2} \le \| \sigma^* \| \le \| \hat{\sigma} \|$,
where $\hat{\sigma}$ is a bounding vector, although there are cases for
which a search is unnecessary because
$\cP(\cC) = \| \sigma^* \| = \|\hat{\sigma}\|$, see~\cite{holder2020}.
The components of $\sigma^*$ do not explicitly provide information about the
original data $\rho^D_{skt}$, $\rho^R_{skt}$, $v_{skt}$, $\ell^D_k$, and 
$\ell^R_k$, and while one could work to back-calculate the uncertainty necessitated
of the original data, especially because doing so could identify influential
political options that might improve a map's standing, we do not
make this effort here and instead base our comments on the transformed
data $X_{skt}$ and $Y_{skt}$.

We reiterate before presenting our results that our goal is not to
advocate our outcomes as a catholicon for the gerrymandering problem. We
only compare eight different maps and make no effort to draw or infer new
or improved district boundaries.  We only use two defendable characteristics
based on sound data but do not consider different, or larger 
collections of, characteristics.  The options and possibilities upon
which an analysis could be made are very, very many, and we
reiterate that our singular goal is to demonstrate that the proximity 
to equitable efficiency is a practical comparison tool that can support 
a data-driven approach to identifying a best congressional map from a 
predetermined collection of maps and an established set of characteristics.

\section{Congressional Map Comparisons} \label{sec-results}

Table~\ref{table-shortList} contains a summary of our comparative study,
and Table 2 in the supplementary material lists all outcomes.  Only maps 
with the smallest proximity values are listed in Table~\ref{table-shortList}, 
and only states with at least five congressional districts are included in
this table.  All states with only two congressional districts had at least 
six maps with proximity values of zero, but all states with at least three 
or four congressional districts had positive minimum proximity values.  This
last observation is somewhat surprising since a low number of DMUs 
would regularly lead to each district having an efficiency score of one.

Most proximity values in Table~\ref{table-shortList}
are less than $0.1$, which suggests that good maps for most states
should have proximity measures below this value.  Some states are much
lower, for instance North Carolina and Washington, and some are higher, 
for instance Pennsylvania.  The compact map for Pennsylvania has a highest
proximity value of $0.2909$, which is $244\%$ of the minimum proximity
value of the Republican map.  This spread suggests substantial disparity
in the districts' efficiency scores for the various maps.  Similarly, the 
ratios of the highest to lowest proximity values for North Carolina and 
Washington are respectively $957\%$ (Compact/Democratic) and $1242\%$ 
(Current/Republican). Another observation is that we rarely invoke the equality
$\cP(\cC) = \| \hat{\sigma} \|$, with the best maps for New Jersey and 
Wisconsin being exceptions. Also notice that Missouri, Alabama, and 
Louisiana have ties for their best map, with the current map being an option 
in each case. The vast majority of states otherwise have minimum proximity
values that are decided by a single map. 

\begin{table}[t!]
\begin{center}
\renewcommand{\arraystretch}{1.1}
{\footnotesize
\begin{tabular}{cccccc}
 & Num. & Most & $\cP(\cC) = \|\hat{\sigma}\|$ & & \\[-1pt]
 & of & Proximal & or & \multicolumn{2}{c}{optimal $\sigma^*$} \\[-1pt]
State & Dist.
      & Map 
      & $\|\hat{\sigma}\|/\sqrt{2} \le \cP(\cC) \le \|\hat{\sigma}\|$
      & $\sigma^*_X$ 
      & $\sigma^*_Y$ \\
\hline \hline
CA & 53 & Rto & $0.0606 \le 0.0857 \le 0.0857$ & $0.0334$ & $0.0789$ \\ 
TX & 36 & Dem & $0.0969 \le 0.0969 \le 0.1371$ & $0.0622$ & $0.0743$ \\ 
FL & 27 & Dem & $0.0870 \le 0.0870 \le 0.1231$ & $0.0607$ & $0.0624$ \\ 
NY & 27 & Rep & $0.0155 \le 0.0219 \le 0.0220$ & $0.0144$ & $0.0165$ \\ 
PA & 18 & Rep & $0.1193 \le 0.1193 \le 0.1688$ & $0.0535$ & $0.1066$ \\ 
IL & 18 & Rep & $0.0813 \le 0.0813 \le 0.1150$ & $0.0644$ & $0.0495$ \\ 
OH & 16 & Rto & $0.0910 \le 0.0910 \le 0.1288$ & $0.0711$ & $0.0568$ \\ 
GA & 14 & Rep & $0.0633 \le 0.0633 \le 0.0896$ & $0.0325$ & $0.0543$ \\ 
MI & 14 & Dem & $0.0640 \le 0.0904 \le 0.0905$ & $0.0775$ & $0.0465$ \\ 
NC & 13 & Dem & $0.0294 \le 0.0294 \le 0.0416$ & $0.0178$ & $0.0234$ \\ 
NJ & 12 & Rep & $0.0762=0.0762$ & $0.0099$ & $0.0755$ \\ 
VA & 11 & Cty & $0.0651 \le 0.0651 \le 0.0921$ & $0.0504$ & $0.0412$ \\ 
WA & 10 & Rep & $0.0136 \le 0.0192 \le 0.0193$ & $0.0183$ & $0.0056$ \\ 
TN & 9 & Rep & $0.0287 \le 0.0287 \le 0.0406$ & $0.0284$ & $0.0034$ \\ 
IN & 9 & Rep & $0.0195 \le 0.0195 \le 0.0276$ & $0.0090$ & $0.0172$ \\ 
MA & 9 & Dem & $0.0317 \le 0.0448 \le 0.0448$ & $0.0244$ & $0.0375$ \\ 
AZ & 9 & Dem & $0.0718 \le 0.1014 \le 0.1015$ & $0.0643$ & $0.0784$ \\ 
WI & 8 & Rep & $0.0988=0.0988$ & $0.0686$ & $0.0710$ \\ 
MD & 8 & Dem & $0.0417 \le 0.0590 \le 0.0590$ & $0.0384$ & $0.0447$ \\ 
MO & 8 & Crt & $0.0448 \le 0.0448 \le 0.0635$ & $0.0422$ & $0.0151$ \\ 
 & & MMn & $0.0448 \le 0.0448 \le 0.0635$ & $0.0422$ & $0.0151$ \\ 
MN & 8 & Dem & $0.0677 \le 0.0677 \le 0.0958$ & $0.0603$ & $0.0306$ \\ 
CO & 7 & Rep & $0.0678 \le 0.0678 \le 0.0960$ & $0.0492$ & $0.0466$ \\ 
AL & 7 & Crt & $0.0067 \le 0.0095 \le 0.0095$ & $0.0095$ & $0.0000$ \\ 
 & & Rep & $0.0067 \le 0.0095 \le 0.0095$ & $0.0095$ & $0.0000$ \\ 
SC & 7 & MMn & $0.0234 \le 0.0234 \le 0.0332$ & $0.0234$ & $0.0000$ \\ 
LA & 6 & Crt & $0.0068 \le 0.0096 \le 0.0097$ & $0.0096$ & $0.0003$ \\ 
 & & Rep & $0.0068 \le 0.0096 \le 0.0097$ & $0.0096$ & $0.0003$ \\ 
KY & 6 & Rep & $0.0393 \le 0.0556 \le 0.0556$ & $0.0360$ & $0.0423$ \\ 
CT & 5 & Dem & $0.0167 \le 0.0167 \le 0.0236$ & $0.0065$ & $0.0153$ \\ 
OK & 5 & Cty & $0.0259 \le 0.0259 \le 0.0367$ & $0.0188$ & $0.0179$ \\ 
OR & 5 & Rep & $0.0052 \le 0.0052 \le 0.0074$ & $0.0038$ & $0.0035$
\end{tabular} }
\vspace*{6pt}
\end{center}
\caption{Summary results for the congressional maps with the
	smallest proximities to equitable efficiency; all 
	states with at least five congressional 
	districts. \hfill} \label{table-shortList}
\end{table}

\begin{figure}[t]
\begin{minipage}[c]{0.48\linewidth}
\includegraphics[width=\linewidth]{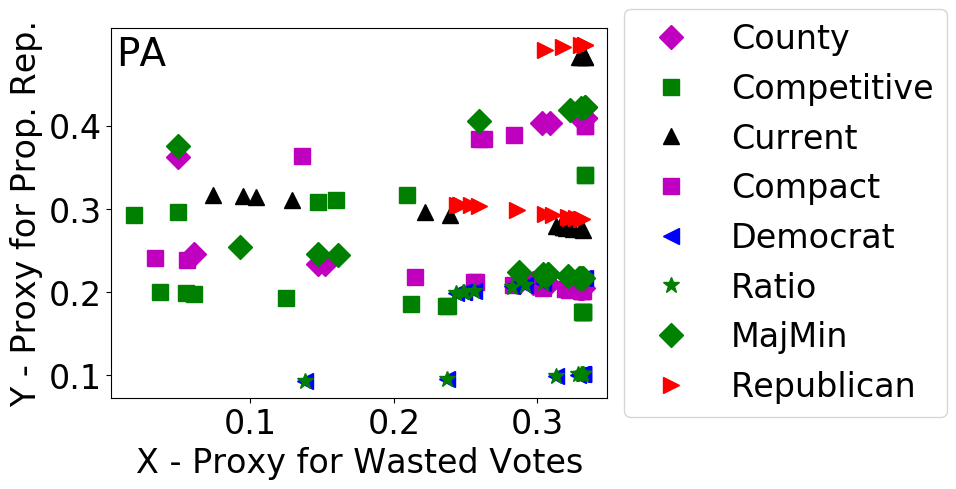}
\end{minipage}
\hfill
\begin{minipage}[c]{0.48\linewidth}
\includegraphics[width=\linewidth]{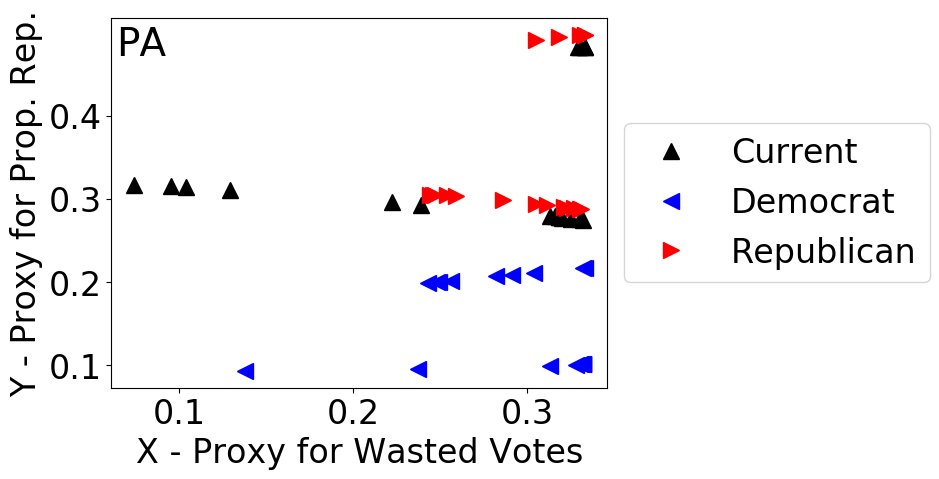}
\end{minipage}
\\\
\begin{minipage}[c]{0.48\linewidth}
\includegraphics[width=\linewidth]{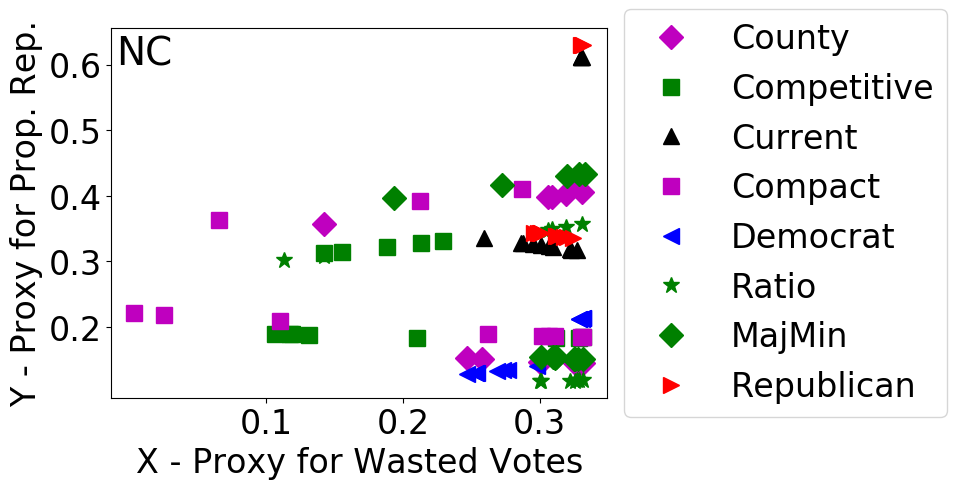}
\end{minipage}
\hfill
\begin{minipage}[c]{0.48\linewidth}
\includegraphics[width=\linewidth]{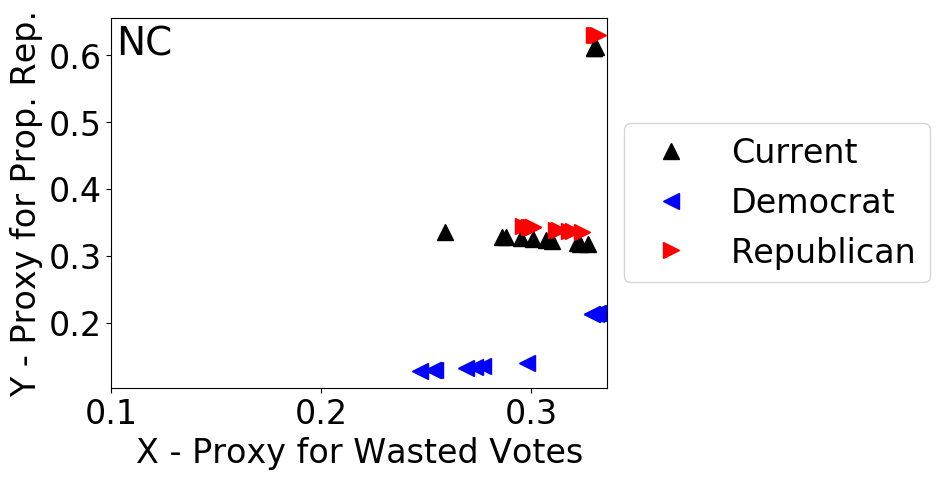}
\end{minipage}
\\\
\begin{minipage}[c]{0.48\linewidth}
\includegraphics[width=\linewidth]{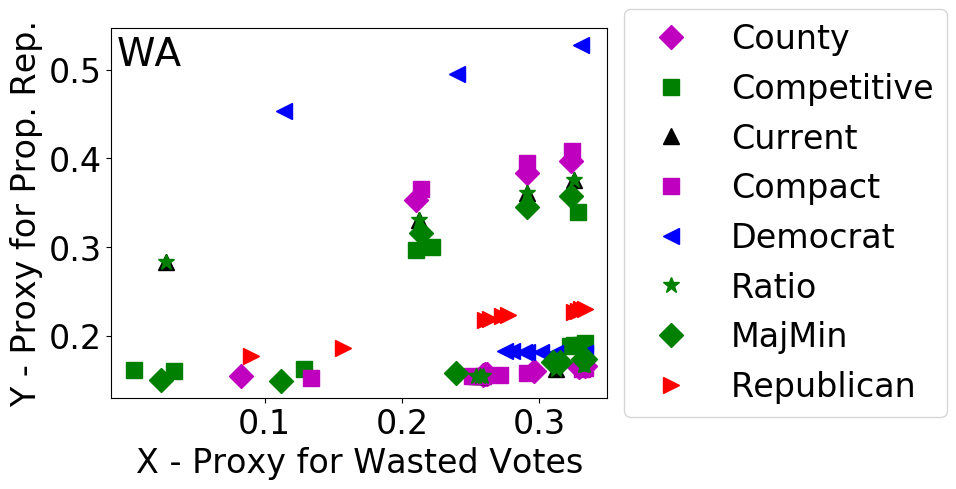}
\end{minipage}
\hfill
\begin{minipage}[c]{0.48\linewidth}
\includegraphics[width=\linewidth]{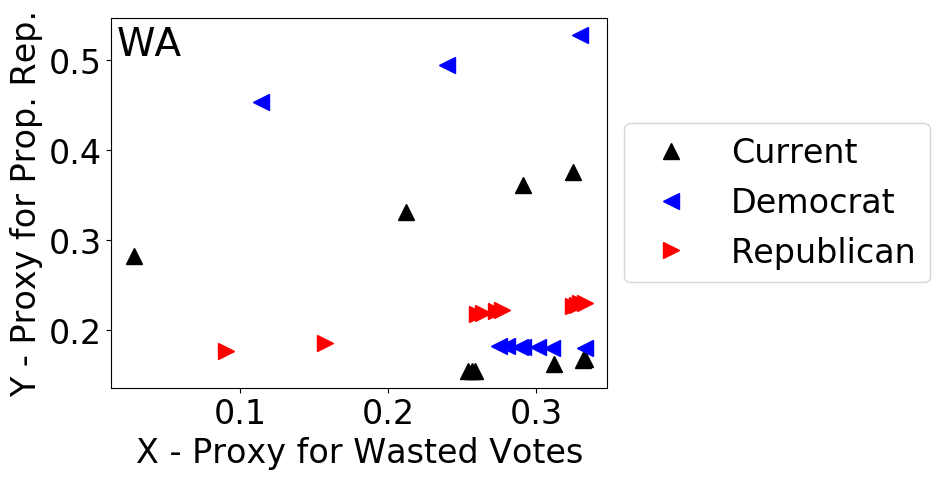}
\end{minipage}
\vspace*{10pt}
\caption{Characteristic data for Pennsylvania, North Carolina, and
	Washington.} \label{fig-charData}
\end{figure}

One way to visualize the disparity of the various maps is to plot their
input and output characteristics against each other, a technique common in 
the DEA literature.  Figure~\ref{fig-charData} depicts the characteristics
for Pennsylvania, North Carolina, and Washington; graphs on the left
show characteristics for all eight maps, and graphs on the right show
characteristics for only the current, Republican, and Democratic maps.
A district with characteristics in the upper left would be favorable
since it would have an anticipated low number of wasted votes and
a high expected value toward achieving proportional representation. 
Districts to the lower right reverse this sentiment and suggest a loss 
in quality as measured by our characteristics.

Consider the graph for North Carolina in Figure~\ref{fig-charData} that
illustrates the current, Republican, and Democratic maps, for which
the proximity measures are respectively $0.1680$, $0.0932$, and $0.0294$.
The Democratic map has the lowest proximity value of the three
even though its districts trend toward the lower right portion of the 
graph, which might seem to suggest that the current and Republican maps 
are better.  However, both the current and Republican maps have districts 
near the top and\hspace*{-0.8pt}/\hspace*{-0.8pt}or left of the graph that 
are somewhat distant from other districts toward the bottom 
and\hspace*{-0.8pt}/\hspace*{-0.8pt}or right, and this spread between their 
best and worst districts increases the value of the proximity to equitable 
efficiency.  The proximity values for the current and Republican maps are 
higher than that of the Democratic map because they require more 
characteristic uncertainty before all districts have an equal claim to 
efficiency with regard to each other.  This fact is important to note: the 
proximity to equitable efficiency is an intra-comparison and not an 
inter-comparison, and a map's proximity to equitable efficiency is 
indifferent to the other maps.  In this case, the districts of the 
Democratic map are more equitable, especially with regard to their role in 
proportional representation, than are the districts of either the current 
or Republican map.  An apt interpretation is that the high quality districts 
of the current and Republican maps force the other districts to have less 
quality with regard to efficiency, and it is this disparity that the 
proximity to equitable efficiency measures.  A similar analysis applies to the 
comparison of the current, Republican, and Democratic maps of Washington,
but in this case the Republican map's proximity value of $0.0192$
bests those of the current and Democratic maps, which are
$0.2385$ and $0.1864$.  A graphical interpretation of Pennsylvania is
less clear, and in this case the proximity values are $0.182$ (current),
$0.1194$ (Republican), and $0.1342$ (Democratic).  Conspicuous
disparities in this case appear, at least primarily, to be the result 
of variations with regard to wasted votes.

\begin{figure}
\begin{center}
\includegraphics[width=0.48\linewidth]{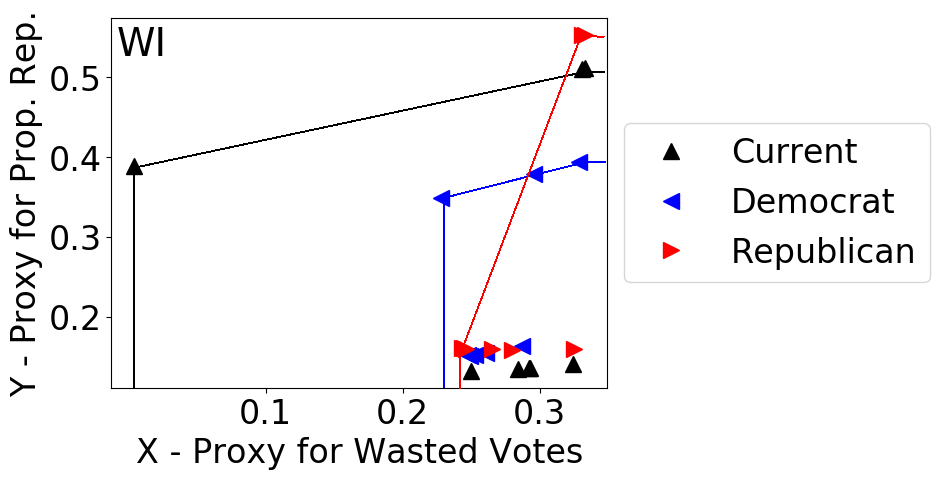}
\end{center}
\caption{Wisconsin's Democratic, Republican, and current maps with
	original efficient frontiers.} \label{fig-nonHomogenous}
\end{figure}

We remind that our proximity measure does not specifically 
favor a map because its districts have more homogeneous characteristics. 
This suspicion would in fact be a misinterpretation of the results just 
presented.  For instance, it is incorrect to assume that the Republican 
map for Washington is favored by our proximity measure simply because its 
data is most clustered.  The Wisconsin map in Figure~\ref{fig-nonHomogenous} 
illustrates this point by adding the traditional efficient frontiers of 
each map.  A map's proximity value is zero only if all districts lie on 
their respective efficient frontier. The Republican map in this case has 
the lowest proximity measure although its data has a greater spread of 
proportional representation and about an equal spread of wasted votes
as compared with the Democratic map.  To exaggerate this point, 
suppose that the current map's districts to the lower right had instead 
appeared on their (black) efficient frontier.  The current map would have 
then had a proximity value of zero, and it would have been preferred over 
both the Democratic and Republican maps even though its districts 
would have had a far greater variation in their characteristics.  Lastly,
our proximity measure is also not a straightforward merger of distances from 
the efficient frontier because the frontier itself depends on how
each district individually selects data from the 
realm of uncertain options.

The results of this section can be recreated with the 
supplemental python code, which is described in the supplementary material.
This code could be altered to replace or add new characteristics or
to assess newly proposed congressional maps. All robust problems were
modeled with Pyomo~\cite{hart2017} and solved with Gurobi.

\section*{Acknowledgements}
The authors are grateful for thoughtful comments made by Ian Ludden 
and Matthias Ehrgott.



\newpage

%


\begin{center}
{\Large Supplmentary Materials for \\[10pt] 
    Comparing  Voting Districts with \\[6pt]
    Uncertain Data Envelopment Analysis} 
\end{center}

\bigskip

\setcounter{section}{0}
\section{Software}

Python code to reproduce the results in {\it Using Uncertain Data 
Envelopment Analysis to Compare Classifications with an Application
in Gerrymandering} is part of the supplementary materials.  Data is 
located in the {\tt data} directory with the following titles: 
\begin{itemize}
\item {\tt CongressionalDistricts.txt} - original source data 
	from FiveThirtyEight's {\it The Atlas of Redistricting},
\item {\tt partyAffiliation.csv} -- original source data from the 
	Pew Research Center,
\item {\tt XdataFile.dat.article} -- transformed data $X_{skt}$,
\item {\tt YdataFile.dat.article} -- transformed data $Y_{skt}$,
\item {\tt gmResultsStateMap.csv} -- results in 
	Table~\ref{table-completeResults} in Section~\ref{sec-table}, and
\item {\tt gmResultsState.csv} -- results for all states like those
	in Table 1 of the article.
\end{itemize}
Python code is in the {\tt ver1} directory, with the primary script being
\begin{center}
{\tt gerrymandering.py}.  
\end{center}
The code runs with Python 2.7.17 and requires Pyomo~\cite{hart2017} and 
Gurobi.  Settings at the top of {\tt gerrymandering.py}
allow a user to set the states and maps to consider, to generate new
data ($Y_{skt}$), to save data, to generate new LaTeX tables, and to
make new data plots like those in Figure 1. New figures
are placed in the {\tt figs} directory in png format.  Completing
the script with the article's data takes several hours, and simulating
new data takes about an hour.

\section{Complete Tabulated Results} \label{sec-table}

Table~\ref{table-completeResults} lists comparative results for
all state and map combinations.  The state order decreases in the
number of congressional districts, and each value of ${\cal P}({\cal C})$
is either bounded $\hat{\sigma}$ or is stated
as an equality.

\begin{center}
\setcounter{table}{1}
\renewcommand{\arraystretch}{1.14}
\begin{longtable}{cccccc}
 & Num. & Most & $\cP(\cC) = \|\hat{\sigma}\|$ & & \\[-2pt]
 & of & Proximal & or & \multicolumn{2}{c}{optimal $\sigma^*$} \\[-2pt]
State & Dist.
      & Map
      & $\|\hat{\sigma}\|/\sqrt{2} \le \cP(\cC) \le \|\hat{\sigma}\|$
      & $\sigma^*_X$
      & $\sigma^*_Y$ \\
\hline \hline
\endhead
CA & 53 & Cty & $0.1152 \le 0.1629 \le 0.1630$ & $0.0866$ & $0.1380$ \\ 
CA & 53 & Cpt & $0.1131 \le 0.1599 \le 0.1600$ & $0.1233$ & $0.1018$ \\ 
CA & 53 & Crt & $0.1361 \le 0.1361 \le 0.1926$ & $0.0992$ & $0.0932$ \\ 
CA & 53 & Cmt & $0.1007 \le 0.1424 \le 0.1424$ & $0.0877$ & $0.1121$ \\ 
CA & 53 & Dem & $0.0954 \le 0.0954 \le 0.1350$ & $0.0669$ & $0.0680$ \\ 
CA & 53 & Rto & $0.0606 \le 0.0857 \le 0.0857$ & $0.0334$ & $0.0789$ \\ 
CA & 53 & MMn & $0.1326 \le 0.1326 \le 0.1876$ & $0.0975$ & $0.0899$ \\ 
CA & 53 & Rep & $0.0697 \le 0.0986 \le 0.0986$ & $0.0510$ & $0.0843$ \\ 
TX & 36 & Cty & $0.1733 \le 0.2451 \le 0.2451$ & $0.1947$ & $0.1489$ \\ 
TX & 36 & Cpt & $0.1369 \le 0.1935 \le 0.1936$ & $0.1371$ & $0.1365$ \\ 
TX & 36 & Crt & $0.1222 \le 0.1222 \le 0.1728$ & $0.0195$ & $0.1206$ \\ 
TX & 36 & Cmt & $0.1751 \le 0.1751 \le 0.2478$ & $0.1125$ & $0.1342$ \\ 
TX & 36 & Dem & $0.0969 \le 0.0969 \le 0.1371$ & $0.0622$ & $0.0743$ \\ 
TX & 36 & Rto & $0.1396 \le 0.1974 \le 0.1975$ & $0.1321$ & $0.1467$ \\ 
TX & 36 & MMn & $0.1314 \le 0.1314 \le 0.1859$ & $0.0632$ & $0.1152$ \\ 
TX & 36 & Rep & $0.1505 \le 0.1505 \le 0.2130$ & $0.1157$ & $0.0962$ \\ 
FL & 27 & Cty & $0.1481 \le 0.1481 \le 0.2095$ & $0.0900$ & $0.1176$ \\ 
FL & 27 & Cpt & $0.1369 \le 0.1369 \le 0.1937$ & $0.0858$ & $0.1067$ \\ 
FL & 27 & Crt & $0.1417 \le 0.2003 \le 0.2004$ & $0.1439$ & $0.1393$ \\ 
FL & 27 & Cmt & $0.1424 \le 0.2013 \le 0.2014$ & $0.1507$ & $0.1334$ \\ 
FL & 27 & Dem & $0.0870 \le 0.0870 \le 0.1231$ & $0.0607$ & $0.0624$ \\ 
FL & 27 & Rto & $0.1198 \le 0.1198 \le 0.1695$ & $-0.0071$ & $0.1196$ \\ 
FL & 27 & MMn & $0.1530 \le 0.2163 \le 0.2163$ & $0.1435$ & $0.1618$ \\ 
FL & 27 & Rep & $0.1515 \le 0.1515 \le 0.2143$ & $0.1029$ & $0.1111$ \\ 
NY & 27 & Cty & $0.1396 \le 0.1974 \le 0.1975$ & $0.1363$ & $0.1427$ \\ 
NY & 27 & Cpt & $0.1618=0.1618$ & $0.1130$ & $0.1157$ \\ 
NY & 27 & Crt & $0.1820 \le 0.2573 \le 0.2573$ & $0.2043$ & $0.1564$ \\ 
NY & 27 & Cmt & $0.1535 \le 0.1535 \le 0.2172$ & $0.0915$ & $0.1232$ \\ 
NY & 27 & Dem & $0.1474 \le 0.2084 \le 0.2085$ & $0.1879$ & $0.0901$ \\ 
NY & 27 & Rto & $0.1723 \le 0.2436 \le 0.2437$ & $0.1587$ & $0.1848$ \\ 
NY & 27 & MMn & $0.1528 \le 0.2160 \le 0.2161$ & $0.1290$ & $0.1733$ \\ 
NY & 27 & Rep & $0.0155 \le 0.0219 \le 0.0220$ & $0.0144$ & $0.0165$ \\ 
PA & 18 & Cty & $0.2129 \le 0.2129 \le 0.3011$ & $0.1349$ & $0.1646$ \\ 
PA & 18 & Cpt & $0.1672 \le 0.2364 \le 0.2365$ & $0.1569$ & $0.1768$ \\ 
PA & 18 & Crt & $0.1824 \le 0.1824 \le 0.2580$ & $0.1378$ & $0.1194$ \\ 
PA & 18 & Cmt & $0.2058 \le 0.2909 \le 0.2910$ & $0.1924$ & $0.2182$ \\ 
PA & 18 & Dem & $0.1342 \le 0.1342 \le 0.1899$ & $0.0968$ & $0.0930$ \\ 
PA & 18 & Rto & $0.1342 \le 0.1342 \le 0.1899$ & $0.0968$ & $0.0930$ \\ 
PA & 18 & MMn & $0.2254 \le 0.2254 \le 0.3188$ & $0.1849$ & $0.1288$ \\ 
PA & 18 & Rep & $0.1193 \le 0.1193 \le 0.1688$ & $0.0535$ & $0.1066$ \\ 
IL & 18 & Cty & $0.0942 \le 0.0942 \le 0.1333$ & $0.0508$ & $0.0793$ \\ 
IL & 18 & Cpt & $0.0981 \le 0.1386 \le 0.1387$ & $0.0905$ & $0.1050$ \\ 
IL & 18 & Crt & $0.1427 \le 0.2017 \le 0.2018$ & $0.1361$ & $0.1488$ \\ 
IL & 18 & Cmt & $0.1032=0.1032$ & $0.0898$ & $0.0507$ \\ 
IL & 18 & Dem & $0.1357 \le 0.1357 \le 0.1920$ & $0.1048$ & $0.0861$ \\ 
IL & 18 & Rto & $0.0974 \le 0.0974 \le 0.1379$ & $0.0781$ & $0.0582$ \\ 
IL & 18 & MMn & $0.1018=0.1018$ & $0.0600$ & $0.0822$ \\ 
IL & 18 & Rep & $0.0813 \le 0.0813 \le 0.1150$ & $0.0644$ & $0.0495$ \\ 
OH & 16 & Cty & $0.2328 \le 0.3291 \le 0.3292$ & $0.2756$ & $0.1798$ \\ 
OH & 16 & Cpt & $0.1685 \le 0.2382 \le 0.2383$ & $0.1709$ & $0.1660$ \\ 
OH & 16 & Crt & $0.1738 \le 0.1738 \le 0.2459$ & $0.0978$ & $0.1436$ \\ 
OH & 16 & Cmt & $0.2593 \le 0.3666 \le 0.3667$ & $0.3194$ & $0.1799$ \\ 
OH & 16 & Dem & $0.1564=0.1564$ & $0.1185$ & $0.1020$ \\ 
OH & 16 & Rto & $0.0910 \le 0.0910 \le 0.1288$ & $0.0711$ & $0.0568$ \\ 
OH & 16 & MMn & $0.2010 \le 0.2842 \le 0.2843$ & $0.2413$ & $0.1500$ \\ 
OH & 16 & Rep & $0.0972 \le 0.1373 \le 0.1374$ & $0.0805$ & $0.1112$ \\ 
GA & 14 & Cty & $0.1654 \le 0.2338 \le 0.2338$ & $0.2208$ & $0.0768$ \\ 
GA & 14 & Cpt & $0.1790 \le 0.2530 \le 0.2531$ & $0.1794$ & $0.1784$ \\ 
GA & 14 & Crt & $0.0880 \le 0.0880 \le 0.1246$ & $0.0849$ & $0.0232$ \\ 
GA & 14 & Cmt & $0.1643 \le 0.2323 \le 0.2323$ & $0.2270$ & $0.0494$ \\ 
GA & 14 & Dem & $0.0686 \le 0.0970 \le 0.0971$ & $0.0670$ & $0.0701$ \\ 
GA & 14 & Rto & $0.1066 \le 0.1507 \le 0.1507$ & $0.1398$ & $0.0561$ \\ 
GA & 14 & MMn & $0.1744 \le 0.2466 \le 0.2467$ & $0.2385$ & $0.0626$ \\ 
GA & 14 & Rep & $0.0633 \le 0.0633 \le 0.0896$ & $0.0325$ & $0.0543$ \\ 
MI & 14 & Cty & $0.2106 \le 0.2978 \le 0.2979$ & $0.2357$ & $0.1820$ \\ 
MI & 14 & Cpt & $0.1864 \le 0.1864 \le 0.2637$ & $0.1123$ & $0.1487$ \\ 
MI & 14 & Crt & $0.1277 \le 0.1277 \le 0.1807$ & $0.1013$ & $0.0777$ \\ 
MI & 14 & Cmt & $0.1878 \le 0.2654 \le 0.2655$ & $0.1847$ & $0.1906$ \\ 
MI & 14 & Dem & $0.0640 \le 0.0904 \le 0.0905$ & $0.0775$ & $0.0465$ \\ 
MI & 14 & Rto & $0.1178=0.1178$ & $0.1024$ & $0.0582$ \\ 
MI & 14 & MMn & $0.1340 \le 0.1340 \le 0.1896$ & $0.1063$ & $0.0816$ \\ 
MI & 14 & Rep & $0.0770 \le 0.1088 \le 0.1089$ & $0.0818$ & $0.0717$ \\ 
NC & 13 & Cty & $0.1647 \le 0.1647 \le 0.2330$ & $0.1632$ & $0.0224$ \\ 
NC & 13 & Cpt & $0.2536=0.2536$ & $0.1680$ & $0.1899$ \\ 
NC & 13 & Crt & $0.1188 \le 0.1680 \le 0.1680$ & $0.1116$ & $0.1255$ \\ 
NC & 13 & Cmt & $0.1991 \le 0.2814 \le 0.2815$ & $0.2180$ & $0.1779$ \\ 
NC & 13 & Dem & $0.0294 \le 0.0294 \le 0.0416$ & $0.0178$ & $0.0234$ \\ 
NC & 13 & Rto & $0.2110 \le 0.2110 \le 0.2984$ & $0.1804$ & $0.1093$ \\ 
NC & 13 & MMn & $0.1530 \le 0.1530 \le 0.2164$ & $0.1372$ & $0.0677$ \\ 
NC & 13 & Rep & $0.0660 \le 0.0932 \le 0.0933$ & $0.0545$ & $0.0756$ \\ 
NJ & 12 & Cty & $0.1961 \le 0.1961 \le 0.2774$ & $0.1537$ & $0.1218$ \\ 
NJ & 12 & Cpt & $0.1328 \le 0.1877 \le 0.1878$ & $0.1509$ & $0.1117$ \\ 
NJ & 12 & Crt & $0.1359=0.1359$ & $0.1038$ & $0.0878$ \\ 
NJ & 12 & Cmt & $0.1612 \le 0.2279 \le 0.2280$ & $0.1583$ & $0.1639$ \\ 
NJ & 12 & Dem & $0.0862 \le 0.1218 \le 0.1219$ & $0.0870$ & $0.0852$ \\ 
NJ & 12 & Rto & $0.1540 \le 0.1540 \le 0.2179$ & $0.1038$ & $0.1138$ \\ 
NJ & 12 & MMn & $0.1237 \le 0.1237 \le 0.1751$ & $0.0827$ & $0.0920$ \\ 
NJ & 12 & Rep & $0.0762=0.0762$ & $0.0099$ & $0.0755$ \\ 
VA & 11 & Cty & $0.0651 \le 0.0651 \le 0.0921$ & $0.0504$ & $0.0412$ \\ 
VA & 11 & Cpt & $0.1883=0.1883$ & $0.1345$ & $0.1317$ \\ 
VA & 11 & Crt & $0.0851 \le 0.0851 \le 0.1205$ & $0.0651$ & $0.0548$ \\ 
VA & 11 & Cmt & $0.1021 \le 0.1021 \le 0.1445$ & $0.0599$ & $0.0827$ \\ 
VA & 11 & Dem & $0.0999 \le 0.1412 \le 0.1413$ & $0.1101$ & $0.0884$ \\ 
VA & 11 & Rto & $0.1112 \le 0.1112 \le 0.1573$ & $0.0811$ & $0.0760$ \\ 
VA & 11 & MMn & $0.0651 \le 0.0920 \le 0.0920$ & $0.0874$ & $0.0285$ \\ 
VA & 11 & Rep & $0.0934 \le 0.1320 \le 0.1321$ & $0.1110$ & $0.0715$ \\ 
WA & 10 & Cty & $0.1546 \le 0.2186 \le 0.2187$ & $0.1396$ & $0.1682$ \\ 
WA & 10 & Cpt & $0.1563 \le 0.1563 \le 0.2212$ & $0.0995$ & $0.1206$ \\ 
WA & 10 & Crt & $0.1687 \le 0.2385 \le 0.2386$ & $0.1772$ & $0.1595$ \\ 
WA & 10 & Cmt & $0.1667 \le 0.2357 \le 0.2357$ & $0.1353$ & $0.1929$ \\ 
WA & 10 & Dem & $0.1864 \le 0.1864 \le 0.2637$ & $0.1776$ & $0.0567$ \\ 
WA & 10 & Rto & $0.1995 \le 0.2820 \le 0.2821$ & $0.2322$ & $0.1601$ \\ 
WA & 10 & MMn & $0.1596 \le 0.1596 \le 0.2258$ & $0.1091$ & $0.1165$ \\ 
WA & 10 & Rep & $0.0136 \le 0.0192 \le 0.0193$ & $0.0183$ & $0.0056$ \\ 
TN & 9 & Cty & $0.1559=0.1559$ & $0.1380$ & $0.0726$ \\ 
TN & 9 & Cpt & $0.1986=0.1986$ & $0.1884$ & $0.0628$ \\ 
TN & 9 & Crt & $0.0826 \le 0.1167 \le 0.1168$ & $0.1167$ & $0.0032$ \\ 
TN & 9 & Cmt & $0.1365 \le 0.1365 \le 0.1930$ & $0.0884$ & $0.1039$ \\ 
TN & 9 & Dem & $0.1061 \le 0.1061 \le 0.1501$ & $0.1061$ & $0.0000$ \\ 
TN & 9 & Rto & $0.2480 \le 0.3506 \le 0.3507$ & $0.3506$ & $0.0000$ \\ 
TN & 9 & MMn & $0.0862 \le 0.1219 \le 0.1219$ & $0.1218$ & $0.0032$ \\ 
TN & 9 & Rep & $0.0287 \le 0.0287 \le 0.0406$ & $0.0284$ & $0.0034$ \\ 
IN & 9 & Cty & $0.0619 \le 0.0619 \le 0.0876$ & $0.0443$ & $0.0432$ \\ 
IN & 9 & Cpt & $0.1825 \le 0.1825 \le 0.2582$ & $0.1716$ & $0.0622$ \\ 
IN & 9 & Crt & $0.1133=0.1133$ & $0.1023$ & $0.0487$ \\ 
IN & 9 & Cmt & $0.1032 \le 0.1032 \le 0.1460$ & $0.1011$ & $0.0205$ \\ 
IN & 9 & Dem & $0.1212 \le 0.1713 \le 0.1714$ & $0.1597$ & $0.0620$ \\ 
IN & 9 & Rto & $0.3227=0.3227$ & $0.2966$ & $0.1271$ \\ 
IN & 9 & MMn & $0.0693 \le 0.0979 \le 0.0980$ & $0.0634$ & $0.0746$ \\ 
IN & 9 & Rep & $0.0195 \le 0.0195 \le 0.0276$ & $0.0090$ & $0.0172$ \\ 
MA & 9 & Cty & $0.0585 \le 0.0585 \le 0.0828$ & $0.0126$ & $0.0571$ \\ 
MA & 9 & Cpt & $0.1758 \le 0.2486 \le 0.2486$ & $0.2479$ & $0.0186$ \\ 
MA & 9 & Crt & $0.0504 \le 0.0711 \le 0.0712$ & $0.0459$ & $0.0543$ \\ 
MA & 9 & Cmt & $0.0684=0.0684$ & $0.0491$ & $0.0475$ \\ 
MA & 9 & Dem & $0.0317 \le 0.0448 \le 0.0448$ & $0.0244$ & $0.0375$ \\ 
MA & 9 & Rto & $0.2899 \le 0.4099 \le 0.4100$ & $0.3522$ & $0.2098$ \\ 
MA & 9 & MMn & $0.0505 \le 0.0713 \le 0.0714$ & $0.0460$ & $0.0544$ \\ 
MA & 9 & Rep & $0.2239=0.2239$ & $0.2190$ & $0.0464$ \\ 
AZ & 9 & Cty & $0.1372 \le 0.1372 \le 0.1941$ & $0.1015$ & $0.0923$ \\ 
AZ & 9 & Cpt & $0.1042 \le 0.1042 \le 0.1474$ & $0.0683$ & $0.0787$ \\ 
AZ & 9 & Crt & $0.1779 \le 0.2515 \le 0.2516$ & $0.1871$ & $0.1680$ \\ 
AZ & 9 & Cmt & $0.2077 \le 0.2937 \le 0.2938$ & $0.2007$ & $0.2143$ \\ 
AZ & 9 & Dem & $0.0718 \le 0.1014 \le 0.1015$ & $0.0643$ & $0.0784$ \\ 
AZ & 9 & Rto & $0.1621 \le 0.2292 \le 0.2293$ & $0.2204$ & $0.0627$ \\ 
AZ & 9 & MMn & $0.1917 \le 0.2711 \le 0.2711$ & $0.2021$ & $0.1807$ \\ 
AZ & 9 & Rep & $0.0806 \le 0.1139 \le 0.1140$ & $0.0750$ & $0.0858$ \\ 
WI & 8 & Cty & $0.1523 \le 0.2154 \le 0.2154$ & $0.1388$ & $0.1647$ \\ 
WI & 8 & Cpt & $0.1833 \le 0.1833 \le 0.2593$ & $0.1262$ & $0.1329$ \\ 
WI & 8 & Crt & $0.2022 \le 0.2858 \le 0.2859$ & $0.2441$ & $0.1487$ \\ 
WI & 8 & Cmt & $0.2013=0.2013$ & $0.1989$ & $0.0309$ \\ 
WI & 8 & Dem & $0.1010 \le 0.1010 \le 0.1429$ & $0.0999$ & $0.0151$ \\ 
WI & 8 & Rto & $0.2270 \le 0.2270 \le 0.3211$ & $0.1537$ & $0.1670$ \\ 
WI & 8 & MMn & $0.1965 \le 0.1965 \le 0.2779$ & $0.1372$ & $0.1406$ \\ 
WI & 8 & Rep & $0.0987=0.0987$ & $0.0686$ & $0.0710$ \\ 
MD & 8 & Cty & $0.2107 \le 0.2980 \le 0.2980$ & $0.2106$ & $0.2108$ \\ 
MD & 8 & Cpt & $0.1637=0.1637$ & $0.0360$ & $0.1597$ \\ 
MD & 8 & Crt & $0.0678 \le 0.0678 \le 0.0959$ & $0.0583$ & $0.0345$ \\ 
MD & 8 & Cmt & $0.1549=0.1549$ & $0.1090$ & $0.1101$ \\ 
MD & 8 & Dem & $0.0417 \le 0.0590 \le 0.0590$ & $0.0384$ & $0.0447$ \\ 
MD & 8 & Rto & $0.1533 \le 0.2167 \le 0.2168$ & $0.1844$ & $0.1138$ \\ 
MD & 8 & MMn & $0.0833=0.0833$ & $0.0524$ & $0.0647$ \\ 
MD & 8 & Rep & $0.0939=0.0939$ & $0.0623$ & $0.0703$ \\ 
MO & 8 & Cty & $0.2369 \le 0.3349 \le 0.3350$ & $0.3241$ & $0.0843$ \\ 
MO & 8 & Cpt & $0.1051 \le 0.1486 \le 0.1487$ & $0.1458$ & $0.0284$ \\ 
MO & 8 & Crt & $0.0448 \le 0.0448 \le 0.0635$ & $0.0422$ & $0.0151$ \\ 
MO & 8 & Cmt & $0.1299=0.1299$ & $0.1195$ & $0.0508$ \\ 
MO & 8 & Dem & $0.0498 \le 0.0703 \le 0.0704$ & $0.0659$ & $0.0245$ \\ 
MO & 8 & Rto & $0.1109 \le 0.1109 \le 0.1569$ & $0.1094$ & $0.0179$ \\ 
MO & 8 & MMn & $0.0448 \le 0.0448 \le 0.0635$ & $0.0422$ & $0.0151$ \\ 
MO & 8 & Rep & $0.1420=0.1420$ & $0.0929$ & $0.1073$ \\ 
MN & 8 & Cty & $0.1455 \le 0.2058 \le 0.2058$ & $0.1448$ & $0.1461$ \\ 
MN & 8 & Cpt & $0.1877 \le 0.2653 \le 0.2654$ & $0.2323$ & $0.1282$ \\ 
MN & 8 & Crt & $0.1706 \le 0.2413 \le 0.2413$ & $0.2304$ & $0.0717$ \\ 
MN & 8 & Cmt & $0.1429 \le 0.2021 \le 0.2021$ & $0.1289$ & $0.1556$ \\ 
MN & 8 & Dem & $0.0677 \le 0.0677 \le 0.0958$ & $0.0603$ & $0.0306$ \\ 
MN & 8 & Rto & $0.1936=0.1936$ & $0.1391$ & $0.1346$ \\ 
MN & 8 & MMn & $0.1790 \le 0.2531 \le 0.2532$ & $0.2320$ & $0.1011$ \\ 
MN & 8 & Rep & $0.1294=0.1294$ & $0.0911$ & $0.0918$ \\ 
CO & 7 & Cty & $0.0984 \le 0.1391 \le 0.1391$ & $0.1157$ & $0.0771$ \\ 
CO & 7 & Cpt & $0.1055 \le 0.1055 \le 0.1492$ & $0.0715$ & $0.0775$ \\ 
CO & 7 & Crt & $0.1015=0.1015$ & $0.0711$ & $0.0724$ \\ 
CO & 7 & Cmt & $0.1013 \le 0.1432 \le 0.1433$ & $0.0987$ & $0.1037$ \\ 
CO & 7 & Dem & $0.1255 \le 0.1255 \le 0.1775$ & $0.1231$ & $0.0245$ \\ 
CO & 7 & Rto & $0.0732 \le 0.1035 \le 0.1035$ & $0.0761$ & $0.0701$ \\ 
CO & 7 & MMn & $0.1114=0.1114$ & $0.1012$ & $0.0467$ \\ 
CO & 7 & Rep & $0.0678 \le 0.0678 \le 0.0960$ & $0.0492$ & $0.0466$ \\ 
AL & 7 & Cty & $0.2479 \le 0.3505 \le 0.3505$ & $0.3504$ & $0.0039$ \\ 
AL & 7 & Cpt & $0.1882 \le 0.2661 \le 0.2662$ & $0.2436$ & $0.1071$ \\ 
AL & 7 & Crt & $0.0067 \le 0.0095 \le 0.0095$ & $0.0095$ & $0.0000$ \\ 
AL & 7 & Cmt & $0.0899 \le 0.1270 \le 0.1271$ & $0.0694$ & $0.1064$ \\ 
AL & 7 & Dem & $0.0590=0.0590$ & $0.0541$ & $0.0234$ \\ 
AL & 7 & Rto & $0.0584=0.0584$ & $0.0531$ & $0.0243$ \\ 
AL & 7 & MMn & $0.0590=0.0590$ & $0.0541$ & $0.0234$ \\ 
AL & 7 & Rep & $0.0067 \le 0.0095 \le 0.0095$ & $0.0095$ & $0.0000$ \\ 
SC & 7 & Cty & $0.2099 \le 0.2967 \le 0.2968$ & $0.2816$ & $0.0934$ \\ 
SC & 7 & Cpt & $0.2909=0.2909$ & $0.2589$ & $0.1326$ \\ 
SC & 7 & Crt & $0.0248=0.0248$ & $0.0245$ & $0.0037$ \\ 
SC & 7 & Cmt & $0.0930 \le 0.0930 \le 0.1316$ & $0.0893$ & $0.0259$ \\ 
SC & 7 & Dem & $0.1672=0.1672$ & $0.1588$ & $0.0524$ \\ 
SC & 7 & Rto & $0.2909=0.2909$ & $0.2589$ & $0.1326$ \\ 
SC & 7 & MMn & $0.0234 \le 0.0234 \le 0.0332$ & $0.0234$ & $0.0000$ \\ 
SC & 7 & Rep & $0.0248=0.0248$ & $0.0245$ & $0.0037$ \\ 
LA & 6 & Cty & $0.1220=0.1220$ & $0.1200$ & $0.0218$ \\ 
LA & 6 & Cpt & $0.1523 \le 0.2154 \le 0.2154$ & $0.2146$ & $0.0180$ \\ 
LA & 6 & Crt & $0.0068 \le 0.0096 \le 0.0097$ & $0.0096$ & $0.0003$ \\ 
LA & 6 & Cmt & $0.0904 \le 0.1278 \le 0.1278$ & $0.0885$ & $0.0921$ \\ 
LA & 6 & Dem & $0.0977 \le 0.0977 \le 0.1382$ & $0.0976$ & $-0.0051$ \\ 
LA & 6 & Rto & $0.0977 \le 0.0977 \le 0.1382$ & $0.0976$ & $-0.0051$ \\ 
LA & 6 & MMn & $0.0957 \le 0.0957 \le 0.1353$ & $0.0953$ & $0.0080$ \\ 
LA & 6 & Rep & $0.0068 \le 0.0096 \le 0.0097$ & $0.0096$ & $0.0003$ \\ 
KY & 6 & Cty & $0.0687=0.0687$ & $0.0601$ & $0.0334$ \\ 
KY & 6 & Cpt & $0.2263 \le 0.2263 \le 0.3201$ & $0.2194$ & $0.0551$ \\ 
KY & 6 & Crt & $0.1064 \le 0.1064 \le 0.1505$ & $0.1062$ & $0.0072$ \\ 
KY & 6 & Cmt & $0.1159=0.1159$ & $0.1093$ & $0.0384$ \\ 
KY & 6 & Dem & $0.0564 \le 0.0797 \le 0.0797$ & $0.0790$ & $0.0105$ \\ 
KY & 6 & Rto & $0.2096 \le 0.2963 \le 0.2964$ & $0.2963$ & $0.0000$ \\ 
KY & 6 & MMn & $0.0941 \le 0.1331 \le 0.1331$ & $0.1179$ & $0.0616$ \\ 
KY & 6 & Rep & $0.0393 \le 0.0556 \le 0.0556$ & $0.0360$ & $0.0423$ \\ 
CT & 5 & Cty & $0.1184 \le 0.1673 \le 0.1674$ & $0.1533$ & $0.0670$ \\ 
CT & 5 & Cpt & $0.1909=0.1909$ & $0.1533$ & $0.1137$ \\ 
CT & 5 & Crt & $0.0360 \le 0.0360 \le 0.0510$ & $0.0053$ & $0.0356$ \\ 
CT & 5 & Cmt & $0.0953 \le 0.1347 \le 0.1348$ & $0.1332$ & $0.0199$ \\ 
CT & 5 & Dem & $0.0167 \le 0.0167 \le 0.0236$ & $0.0065$ & $0.0153$ \\ 
CT & 5 & Rto & $0.2389 \le 0.3377 \le 0.3378$ & $0.3066$ & $0.1415$ \\ 
CT & 5 & MMn & $0.1512 \le 0.2137 \le 0.2138$ & $0.1945$ & $0.0886$ \\ 
CT & 5 & Rep & $0.2149=0.2149$ & $0.1519$ & $0.1519$ \\ 
OK & 5 & Cty & $0.0259 \le 0.0259 \le 0.0367$ & $0.0188$ & $0.0179$ \\ 
OK & 5 & Cpt & $0.0863 \le 0.1220 \le 0.1221$ & $0.0673$ & $0.1017$ \\ 
OK & 5 & Crt & $0.0831=0.0831$ & $0.0565$ & $0.0609$ \\ 
OK & 5 & Cmt & $0.0825=0.0825$ & $0.0506$ & $0.0652$ \\ 
OK & 5 & Dem & $0.1683 \le 0.2379 \le 0.2380$ & $0.2035$ & $0.1232$ \\ 
OK & 5 & Rto & $0.2245=0.2245$ & $0.2035$ & $0.0946$ \\ 
OK & 5 & MMn & $0.0968 \le 0.1368 \le 0.1369$ & $0.0914$ & $0.1018$ \\ 
OK & 5 & Rep & $0.0831=0.0831$ & $0.0565$ & $0.0609$ \\ 
OR & 5 & Cty & $0.1007=0.1007$ & $0.0700$ & $0.0723$ \\ 
OR & 5 & Cpt & $0.0654 \le 0.0654 \le 0.0925$ & $0.0134$ & $0.0640$ \\ 
OR & 5 & Crt & $0.1131=0.1131$ & $0.0798$ & $0.0801$ \\ 
OR & 5 & Cmt & $0.0648 \le 0.0648 \le 0.0918$ & $0.0043$ & $0.0647$ \\ 
OR & 5 & Dem & $0.1198=0.1198$ & $0.0848$ & $0.0845$ \\ 
OR & 5 & Rto & $0.1250=0.1250$ & $0.0761$ & $0.0992$ \\ 
OR & 5 & MMn & $0.0687=0.0687$ & $0.0658$ & $0.0197$ \\ 
OR & 5 & Rep & $0.0052 \le 0.0052 \le 0.0074$ & $0.0038$ & $0.0035$ \\ 
NV & 4 & Cty & $0.1745=0.1745$ & $0.0587$ & $0.1643$ \\ 
NV & 4 & Cpt & $0.0294=0.0294$ & $0.0167$ & $0.0242$ \\ 
NV & 4 & Crt & $0.1009=0.1009$ & $0.0712$ & $0.0715$ \\ 
NV & 4 & Cmt & $0.1069=0.1069$ & $0.0143$ & $0.1060$ \\ 
NV & 4 & Dem & $0.0358=0.0358$ & $0.0358$ & $0.0000$ \\ 
NV & 4 & Rto & $0.1094=0.1094$ & $0.0848$ & $0.0690$ \\ 
NV & 4 & MMn & $0.0193 \le 0.0193 \le 0.0273$ & $0.0193$ & $0.0000$ \\ 
NV & 4 & Rep & $0.1433=0.1433$ & $0.1397$ & $0.0319$ \\ 
AR & 4 & Cty & $0.0285=0.0285$ & $0.0007$ & $0.0285$ \\ 
AR & 4 & Cpt & $0.1778 \le 0.2514 \le 0.2515$ & $0.2509$ & $0.0162$ \\ 
AR & 4 & Crt & $0.0345 \le 0.0345 \le 0.0489$ & $0.0307$ & $0.0158$ \\ 
AR & 4 & Cmt & $0.0447=0.0447$ & $0.0304$ & $0.0327$ \\ 
AR & 4 & Dem & $0.1160 \le 0.1160 \le 0.1642$ & $0.1155$ & $0.0108$ \\ 
AR & 4 & Rto & $0.1226 \le 0.1226 \le 0.1734$ & $0.1163$ & $0.0386$ \\ 
AR & 4 & MMn & $0.0362 \le 0.0362 \le 0.0513$ & $0.0362$ & $0.0000$ \\ 
AR & 4 & Rep & $0.0345 \le 0.0345 \le 0.0489$ & $0.0307$ & $0.0158$ \\ 
IA & 4 & Cty & $0.0383 \le 0.0541 \le 0.0542$ & $0.0466$ & $0.0275$ \\ 
IA & 4 & Cpt & $0.0359 \le 0.0507 \le 0.0507$ & $0.0455$ & $0.0223$ \\ 
IA & 4 & Crt & $0.1002=0.1002$ & $0.0819$ & $0.0576$ \\ 
IA & 4 & Cmt & $0.0933=0.0933$ & $0.0820$ & $0.0445$ \\ 
IA & 4 & Dem & $0.1433=0.1433$ & $0.1013$ & $0.1013$ \\ 
IA & 4 & Rto & $0.0359 \le 0.0507 \le 0.0507$ & $0.0455$ & $0.0223$ \\ 
IA & 4 & MMn & $0.1658 \le 0.2344 \le 0.2344$ & $0.2337$ & $0.0183$ \\ 
IA & 4 & Rep & $0.0553 \le 0.0781 \le 0.0782$ & $0.0557$ & $0.0547$ \\ 
UT & 4 & Cty & $0.1011=0.1011$ & $0.1008$ & $0.0077$ \\ 
UT & 4 & Cpt & $0.1011=0.1011$ & $0.1008$ & $0.0077$ \\ 
UT & 4 & Crt & $0.0052 \le 0.0052 \le 0.0074$ & $0.0031$ & $0.0041$ \\ 
UT & 4 & Cmt & $0.1679=0.1679$ & $0.1230$ & $0.1142$ \\ 
UT & 4 & Dem & $0.0695=0.0695$ & $0.0613$ & $0.0326$ \\ 
UT & 4 & Rto & $0.0695=0.0695$ & $0.0613$ & $0.0326$ \\ 
UT & 4 & MMn & $0.0887=0.0887$ & $0.0852$ & $0.0246$ \\ 
UT & 4 & Rep & $0.0052 \le 0.0052 \le 0.0074$ & $0.0031$ & $0.0041$ \\ 
KS & 4 & Cty & $0.0349 \le 0.0349 \le 0.0494$ & $0.0046$ & $0.0346$ \\ 
KS & 4 & Cpt & $0.0109 \le 0.0154 \le 0.0155$ & $0.0099$ & $0.0118$ \\ 
KS & 4 & Crt & $0.0386 \le 0.0545 \le 0.0546$ & $0.0345$ & $0.0422$ \\ 
KS & 4 & Cmt & $0.0412 \le 0.0583 \le 0.0583$ & $0.0394$ & $0.0428$ \\ 
KS & 4 & Dem & $0.0878 \le 0.1242 \le 0.1242$ & $0.0964$ & $0.0782$ \\ 
KS & 4 & Rto & $0.1763 \le 0.1763 \le 0.2494$ & $0.1762$ & $0.0055$ \\ 
KS & 4 & MMn & $0.1472 \le 0.2081 \le 0.2082$ & $0.2080$ & $0.0063$ \\ 
KS & 4 & Rep & $0.0686=0.0686$ & $0.0622$ & $0.0287$ \\ 
MS & 4 & Cty & $0.1141 \le 0.1613 \le 0.1614$ & $0.1568$ & $0.0379$ \\ 
MS & 4 & Cpt & $0.1617 \le 0.2287 \le 0.2287$ & $0.2286$ & $0.0035$ \\ 
MS & 4 & Crt & $0.0072 \le 0.0101 \le 0.0102$ & $0.0099$ & $0.0022$ \\ 
MS & 4 & Cmt & $0.1276=0.1276$ & $0.1255$ & $0.0229$ \\ 
MS & 4 & Dem & $0.0976=0.0976$ & $0.0976$ & $0.0000$ \\ 
MS & 4 & Rto & $0.1617 \le 0.2287 \le 0.2287$ & $0.2286$ & $0.0035$ \\ 
MS & 4 & MMn & $0.0853 \le 0.1206 \le 0.1206$ & $0.0594$ & $0.1049$ \\ 
MS & 4 & Rep & $0.0072 \le 0.0101 \le 0.0102$ & $0.0099$ & $0.0022$ \\ 
WV & 3 & Cty & $0.0070=0.0070$ & $0.0043$ & $0.0055$ \\ 
WV & 3 & Cpt & $0.0033 \le 0.0033 \le 0.0048$ & $0.0000$ & $0.0033$ \\ 
WV & 3 & Crt & $0.0049 \le 0.0049 \le 0.0070$ & $0.0027$ & $0.0041$ \\ 
WV & 3 & Cmt & $0.0067=0.0067$ & $0.0060$ & $0.0030$ \\ 
WV & 3 & Dem & $0.0033 \le 0.0033 \le 0.0048$ & $0.0000$ & $0.0033$ \\ 
WV & 3 & Rto & $0.0033 \le 0.0033 \le 0.0048$ & $0.0000$ & $0.0033$ \\ 
WV & 3 & MMn & $0.0033 \le 0.0033 \le 0.0048$ & $0.0000$ & $0.0033$ \\ 
WV & 3 & Rep & $0.0049 \le 0.0049 \le 0.0070$ & $0.0027$ & $0.0041$ \\ 
NM & 3 & Cty & $0.0927=0.0927$ & $0.0025$ & $0.0927$ \\ 
NM & 3 & Cpt & $0.0000=0.0000$ & $0.0000$ & $0.0000$ \\ 
NM & 3 & Crt & $0.0000=0.0000$ & $0.0000$ & $0.0000$ \\ 
NM & 3 & Cmt & $0.0000=0.0000$ & $0.0000$ & $0.0000$ \\ 
NM & 3 & Dem & $0.0000=0.0000$ & $0.0000$ & $0.0000$ \\ 
NM & 3 & Rto & $0.0000=0.0000$ & $0.0000$ & $0.0000$ \\ 
NM & 3 & MMn & $0.0035=0.0035$ & $0.0035$ & $0.0000$ \\ 
NM & 3 & Rep & $0.0196=0.0196$ & $0.0139$ & $0.0139$ \\ 
NE & 3 & Cty & $0.0070=0.0070$ & $0.0050$ & $0.0049$ \\ 
NE & 3 & Cpt & $0.0064=0.0064$ & $0.0024$ & $0.0060$ \\ 
NE & 3 & Crt & $0.0070=0.0070$ & $0.0055$ & $0.0042$ \\ 
NE & 3 & Cmt & $0.0046=0.0046$ & $0.0000$ & $0.0046$ \\ 
NE & 3 & Dem & $0.0231 \le 0.0326 \le 0.0326$ & $0.0214$ & $0.0245$ \\ 
NE & 3 & Rto & $0.0231 \le 0.0326 \le 0.0326$ & $0.0214$ & $0.0245$ \\ 
NE & 3 & MMn & $0.0050=0.0050$ & $0.0000$ & $0.0050$ \\ 
NE & 3 & Rep & $0.0028 \le 0.0028 \le 0.0040$ & $0.0009$ & $0.0026$ \\ 
HI & 2 & Cty & $0.0000=0.0000$ & $0.0000$ & $0.0000$ \\ 
HI & 2 & Cpt & $0.0000=0.0000$ & $0.0000$ & $0.0000$ \\ 
HI & 2 & Crt & $0.0000=0.0000$ & $0.0000$ & $0.0000$ \\ 
HI & 2 & Cmt & $0.0000=0.0000$ & $0.0000$ & $0.0000$ \\ 
HI & 2 & Dem & $0.0000=0.0000$ & $0.0000$ & $0.0000$ \\ 
HI & 2 & Rto & $0.0000=0.0000$ & $0.0000$ & $0.0000$ \\ 
HI & 2 & MMn & $0.0000=0.0000$ & $0.0000$ & $0.0000$ \\ 
HI & 2 & Rep & $0.0000=0.0000$ & $0.0000$ & $0.0000$ \\ 
NH & 2 & Cty & $0.0099=0.0099$ & $0.0099$ & $0.0000$ \\ 
NH & 2 & Cpt & $0.0000=0.0000$ & $0.0000$ & $0.0000$ \\ 
NH & 2 & Crt & $0.0000=0.0000$ & $0.0000$ & $0.0000$ \\ 
NH & 2 & Cmt & $0.0215=0.0215$ & $0.0152$ & $0.0152$ \\ 
NH & 2 & Dem & $0.0000=0.0000$ & $0.0000$ & $0.0000$ \\ 
NH & 2 & Rto & $0.0000=0.0000$ & $0.0000$ & $0.0000$ \\ 
NH & 2 & MMn & $0.0000=0.0000$ & $0.0000$ & $0.0000$ \\ 
NH & 2 & Rep & $0.0000=0.0000$ & $0.0000$ & $0.0000$ \\ 
ID & 2 & Cty & $0.0000=0.0000$ & $0.0000$ & $0.0000$ \\ 
ID & 2 & Cpt & $0.0000=0.0000$ & $0.0000$ & $0.0000$ \\ 
ID & 2 & Crt & $0.0000=0.0000$ & $0.0000$ & $0.0000$ \\ 
ID & 2 & Cmt & $0.0000=0.0000$ & $0.0000$ & $0.0000$ \\ 
ID & 2 & Dem & $0.0000=0.0000$ & $0.0000$ & $0.0000$ \\ 
ID & 2 & Rto & $0.0000=0.0000$ & $0.0000$ & $0.0000$ \\ 
ID & 2 & MMn & $0.0000=0.0000$ & $0.0000$ & $0.0000$ \\ 
ID & 2 & Rep & $0.0000=0.0000$ & $0.0000$ & $0.0000$ \\ 
ME & 2 & Cty & $0.0000=0.0000$ & $0.0000$ & $0.0000$ \\ 
ME & 2 & Cpt & $0.0000=0.0000$ & $0.0000$ & $0.0000$ \\ 
ME & 2 & Crt & $0.0000=0.0000$ & $0.0000$ & $0.0000$ \\ 
ME & 2 & Cmt & $0.0000=0.0000$ & $0.0000$ & $0.0000$ \\ 
ME & 2 & Dem & $0.0000=0.0000$ & $0.0000$ & $0.0000$ \\ 
ME & 2 & Rto & $0.0000=0.0000$ & $0.0000$ & $0.0000$ \\ 
ME & 2 & MMn & $0.0000=0.0000$ & $0.0000$ & $0.0000$ \\ 
ME & 2 & Rep & $0.0000=0.0000$ & $0.0000$ & $0.0000$ \\ 
RI & 2 & Cty & $0.0000=0.0000$ & $0.0000$ & $0.0000$ \\ 
RI & 2 & Cpt & $0.0000=0.0000$ & $0.0000$ & $0.0000$ \\ 
RI & 2 & Crt & $0.0000=0.0000$ & $0.0000$ & $0.0000$ \\ 
RI & 2 & Cmt & $0.0000=0.0000$ & $0.0000$ & $0.0000$ \\ 
RI & 2 & Dem & $0.0000=0.0000$ & $0.0000$ & $0.0000$ \\ 
RI & 2 & Rto & $0.0000=0.0000$ & $0.0000$ & $0.0000$ \\ 
RI & 2 & MMn & $0.0000=0.0000$ & $0.0000$ & $0.0000$ \\ 
RI & 2 & Rep & $0.0000=0.0000$ & $0.0000$ & $0.0000$ \\ \\
\\[-20pt]
\caption{Complete results listing all state and map combinations.}
	\label{table-completeResults}
\end{longtable}
\end{center}


\end{document}